# Hydrolytic degradation of ROMP thermosetting materials catalysed by bio-derived acids and enzyme: from network to linear materials

S. Hou,[a] D. M. Hoyle,[b]* C. J. Blackwell,[a] K. Haernvall,[c] V. Perz,[c] G. M. Guebitz[d] and E. Khosravi[a]†

(a) Department of Chemistry, Durham University, Durham, DH1 3LE

(b) Department of Physics, Durham University, Durham, DH1 3LE

(c) Austrian Centre of Industrial Biotechnology GmbH, Konrad Lorenz Strasse 20, 3430 Tulln an der Donau, Austria.

(d) University of Natural Resources and Life Sciences, Institute for Environmental Biotechnology, Konrad Lorenz Strasse 20, 3430 Tulln an der Donau, Austria.

† Corresponding author: E. Khosravi, ezat.khosravi@durham.ac.uk

* Corresponding author: D. M. Hoyle, d.m.hoyle@durham.ac.uk

This paper reports the first example of the degradable ROMP thermosetting materials catalysed by bio-derived acids and cutinase from *Thermobifida cellulosilytica* (Thc_Cut1). The ROMP thermosetting materials are based on norbornene dicarboximides containing acetal ester groups only in the crosslink moiety. The insoluble cross-linked materials were subjected to acid-catalysed hydrolysis using bio-derived acetic and citric acids as well as enzymatic degradation using Thc_Cut1, resulting in the materials becoming completely soluble in dichloromethane. $^1$H NMR and rheological analysis performed on materials after acid-catalysed hydrolysis showed characteristics indistinguishable to that of the linear polymer analogues. These analyses confirmed the cleavage of the crosslink moiety upon degradation with the main backbone chains remaining intact. The glass transition temperatures of the polymer materials after acid-catalysed hydrolysis were the same of those observed for the linear polymer analogue. TGA showed that the cross-linked polymers thermally stable to 150 °C, beyond which showed weight losses due to the thermal cleavage of the acetal ester linkages.

## Introduction

Thermosets are an important class of materials due to their good adhesive strength and high temperature stability.[1,2] However, they are known to be non-degradable, non-reworkable and non-recyclable which is of great significance for both environmental and economic considerations.

Several research groups have reported thermosetting materials that undergo chemical and thermal degradation. This has been achieved by incorporating suitable degradable functional groups into the polymer chains at their design stage. Epoxy resins cured with aromatic diamines containing disulfide linkages have been prepared and shown to degrade under reductive conditions.[3,4] Furthermore, epoxy resins containing acetal linkages in the polymer backbone have been shown to degrade into small soluble fragments when subjected to acid catalysed hydrolysis using phosphoric acid[5] and hydrochloric acid.[6] Linear poly(amidoamine)s containing acetal or ketal groups in the polymer backbone have also been reported to degrade in buffer solution at pH 5 affording low molecular weight and soluble fragments.[7] Furthermore, linear poly(1,4-phenylene dimethylene ketal) which contains ketal linkages in its backbone has been subjected to acid catalysed hydrolysis at pH 1.0 to 7.4 using buffer solutions, yielding low molecular weight fragments.[8] Notably, they reported that after 72 h of acid-catalysed hydrolysis only the polymer at pH 1.0 was 100% hydrolysed whereas at higher pH the polymer was only partially hydrolysed. Moreover, hyperbranched polyglycerols with randomly distributed ketal groups have been shown to undergo degradation at pH 1.1 to 8.2 yielding low molecular weight products.[9] However, it should be noted that the acid-catalysed processes used in these reports lead to complete degradation of the materials.

Cutinase from the compost organism *Thermobifida cellulosilytica* (Thc_Cut1) has been proven to degrade a variety of synthetic polyesters using mild and environmentally friendly conditions.[10,11,12]

Ring opening metathesis polymerisation (ROMP) initiated by Grubbs well-defined ruthenium initiators has been utilised to synthesise well-defined polymers with controlled architectures, molecular weights, dispersities and terminal functionalities.[13] Thermosetting materials have been prepared from ROMP of mixtures of mono- and di-functional norbornene dicarboximide monomers, both containing acetal ester groups.[14,15] They have been shown to be thermally degradable around 200 °C, which is in the same region as the lead-free solder reflow temperature (217 °C); advantageous in the electronic industry as it would allow easy chip removal and replacement, if these materials are used as adhesives. However, but it was difficult to accurately correlate the weight loss, upon heating, of material to the breakdown of acetal ester within the crosslink moiety.

In the work presented here, a range of well-defined thermosetting ROMP materials were prepared with the incorporation of acetal ester groups only in the crosslink moiety. This was achieved through ROMP of a mixture of mono-functional *N*-alkyl norbornene dicarboximides monomer and acetal ester containing di-functional *N*-alkyl

bis(norbornene dicarboximides). A wide range of thermosetting materials with different crosslinking densities were prepared by varying the amounts of difunctional content.

The crucial part in the design of these materials was the cleavage of only the acetal ester linkages via bio-derived acid-catalysed hydrolysis and enzymatic degradation, leaving the main chains intact. This will allow, at low temperature, firstly, the transition from thermosetting to thermoplastic, increasing the recyclability of the material and secondly, developing industrially important class of adhesives that are detachable/re-workable. Degradation behaviour was monitored using %mass loss, NMR, rheological GC-MS analysis. To the best of our knowledge this is the first example of degradable ROMP cross-linked materials via bio-derived acid-catalysed hydrolysis and enzymatic degradation using cutinase (Thc_Cut1).

## Experimental

### Materials

4-Phenylbutylamine and Grubbs 1st generation ruthenium initiator was purchased from Sigma Aldrich and used without further purification. Dichloromethane (DCM) (Analytical Grade, Fisher Scientific), hexane (Analytical Grade, Fisher Scientific), chloroform (HPLC grade, 99.5%, Fisher Scientific), acetone (Analytical Grade, Fisher Scientific), ethyl acetate (Analytical Grade, Fisher Scientific) were used as supplied. Dry toluene and dry DCM were acquired from the departmental solvent purification system. All glassware was oven dried and purged with nitrogen prior to use. All polymerisations were carried out under inert atmosphere.

### Measurements

NMR spectra were either recorded on a Bruker Avance 400 spectrometer at 400.0 MHz ($^1$H) and 100.6 MHz ($^{13}$C); all chemical shifts were referenced to the residual proton impurity of the deuterated solvent, $CDCl_3$. Infrared spectra were recorded using a Perkin Elmer RX1 FT-IR machine.

The gel fraction contents of the cross-linked polymers were determined by sol-gel extraction in DCM and thoroughly dried in an oven at 50 °C for 24 h under reduced pressure. The gel fraction content (%) was determined by the final polymer weight after extraction and drying, divided by the initial weight before extraction.

Thermogravimetric analysis (TGA) measurements were carried performed using a Perkin Elmer Pyris 1 under Helium atmosphere rom 25-400 °C at a heating rate of 10 °C min$^{-1}$.

Differential Scanning Calorimeter measurements were performed on a TA-Q1000 Instrument purged with helium and run -50-150 °C at a heating rate of 10 °C min$^{-1}$.

The products of enzymatic degradation of cross-linked polymers **6-8** were investigated by GC-MS. The hydrolysed samples were extracted with DCM (1:1 volume). For derivatisation of the expected released alcohol molecules, the hydrolysed samples were freeze dried and dissolved in BSTFA reagent mixture (1 mL) (CAS Number: 25561-30-2, Sigma). The closed vial was then heated at 70 °C for 2 h and diluted with DCM before GC-MS analysis.

Samples were analysed by GC-MS (Agilent 7890A GC system, CT analytics CombiPAL and Agilent 5975C VL MSD with triple-Axis detector) under the following conditions. Split less 1 µl injections were made onto a DB-17 MS column (30 m*250 µm*0.25 µm) at a constant flow rate of 1.2 mL min$^{-1}$. The GC oven was operated from 50 °C to 340 °C; 50 °C for 2 min with an increase of 7 °C min$^{-1}$ to 100 °C, kept constant for 1 min, increase of 6 °C min$^{-1}$ to 220 °C, kept constant for 2 min, increase of 20 °C min$^{-1}$ to 340 °C, kept constant for 5 min for a total run time of 43.143 min. The transfer line and injector were set at 300 °C. The detector temperatures were 230 °C for the source and 150 °C for the quadrupole.

### Expression and purification of cutinase

Cutinase (Thc_Cut1) was expressed in E. coli BL21-Gold(DE3) cells which has previously been reported by Acero et al.[11] Freshly transformed E.coli BL21-Gold(DE3) cells were incubated overnight in LB-medium supplemented with kanamycin (40 µg ml-1) on a rotary shaker at 37 °C and 130 rpm. A main culture (400 mL) based on LB-medium supplemented with kanamycin (40 µg mL-1) was inoculated with the overnight culture to an OD of 0.1 and incubated in a rotary shaker at 37 °C and 140 rpm. When the culture reached an optical density (600 nm) of 0.6, isothiopropyl-β-D-galactoside (IPTG) was used to a final concentration of 0.05 mM to induce the expression. The culture was incubated for another 21 h at 20 °C and 130 rpm before the cells were harvested by centrifugation (4500 g, 4 °C, 20 min, Sorvall RC-5B Refrigerated Superspeed Centrifuge, Du Pont Instruments, USA). The pellets were re-suspended in lysis buffer [20 mM NaH2PO4, 500 mM NaCl and 10 mM imidazol, pH 7.4. Sonication (RANSON Ultrasonics cell disruptor USA)] for 6 minutes with a cycle of 1 second sonication and 4 second pause with amplitude of 60 % were used to disrupt the cells. The cell debris were removed by centrifugation (60 minutes, 18,000 rpm, 4 °C) (Sorvall RC-5B Refrigerated Superspeed Centrifuge, Du Pont Instruments, USA) before the enzyme purification. To enable enzyme purification a 6xHis peptide was C-terminally fused over an Ala-Leu-Glu linker sequence to Thc_Cut1. Thc_Cut1 was purified from the supernatant by immobilised metal ion affinity chromatography (IMAC) using HisTrap FF 5 ml columns coupled with ÄKTA purifier 900 (GE Healthcare, UK). The column was equilibrated with lysis buffer before supernatant (30 mL) were loaded on the column. The purification was performed with a buffer (500 mM NaCl, 20 mM NaH2PO4, pH 7.4) to elute the enzyme where the imidazole concentration increased from 10 – 500 mM during 10CV. Fractions containing Thc_Cut1 were pooled and concentrated by centrifugation with Vivaspin 20 column (MWCO of 10,000 Da) (Sartorius AG, Germany). The elution buffer was exchanged with TRIS-HCl (100 mM, pH 7.0) by the use of PD-10 desalting columns (Amersham Biosciences).

Bradford based Bio-99 Rad Protein Assay (Bio-Rad Laboratories GmbH, Munich, Germany) was performed according to the manufacturers' instruction to determine the protein concentration of the purified enzymes. Bovine serum albumin (BSA) was used as standard. SDS-PAGE analysis was performed corresponding to Laemmli proteins stained with Coomassie Brilliant Blue R-250.[16]

## Synthesis of monomers

*Exo*-norbornene dicarboxylic anhydride was prepared following previous published literature.[14]

Monomer **1** was synthesised via a Diels-Alder reaction shown in scheme 1. *Exo*-norbornene dicarboxylic anhydride (4 g, 24.37 mmol), 4-Phenylbutylamine (4 g, 26.81 mmol) and a stirrer bar were placed in a two necked round bottom flask fitted with a rubber septum and a reflux condenser. Dry toluene (16 mL) was added to the flask by a syringe and the mixture was heated to reflux (115 °C) for 17 h. The reaction mixture was left to cool to ambient temperature and toluene was removed under reduced pressure yielding a colourless viscous liquid, **1** (6.41 g, 21.69 mmol, 89% yield). $\delta_H$ (500 MHz, CDCl$_3$) 7.41 (5H, m, Ar-H), 6.33 (2H, s, =CH), 3.39 (2H, s, N-CH$_2$), 3.32 (2H, d, CH), 2.84 (2H, s, CH), 1.60 (1H, m, CH$_2$), 1.48 (1H, d, $J_{HH}$ = 10.8 Hz, CH$_2$), 1.32 (6H, m, CH$_2$). $\delta_C$ (126 MHz, CDCl$_3$) 177.5, 151.9, 138.2, 126.5, 48.1, 46.1, 43.2, 35.0, 28.2, 22.8.

Monomer **2** was similarly synthesised and purified. This involved a Diels-Alder reaction of *Exo*-norbornene dicarboxylic acid (10.00 g, 60.92 mmol) and 2-ethyl-1-hexylamine (8.66 g, 67.01 mmol) resulting in a colourless viscous liquid, **2** (15.94 g, 57.87 mmol, 95% yield). Elemental Anal. Calcd for C$_{17}$H$_{25}$NO$_2$: C, 74.14; H, 9.15; N, 5.09; Found C, 74.11; H, 9.14; N, 5.08. δ (500MHz, CDCl$_3$) 6.12 (2H, s, =CH), 3.17 (2H, d, N-CH$_2$), 3.09 (2H, s, CH), 2.51 (2H, s, CH), 1.52 (1H, m, CH$_2$), 1.32 (1H, d, $J_{HH}$= 10.9, CH$_2$), 1.09 (9H, m, CH$_2$ and CH), 0.71 (6H, t, CH$_3$). $\delta_C$ (126 MHz, CDCl$_3$): 178.0, 137.6, 47.6, 45.0, 42.6, 42.2, 37.6, 30.3, 28.2, 23.7, 22.8, 13.9, 10.2.

The detailed assignment of the $^1$H and $^{13}$C resonances for monomers **1** and **2** can be found in the Supporting Information.

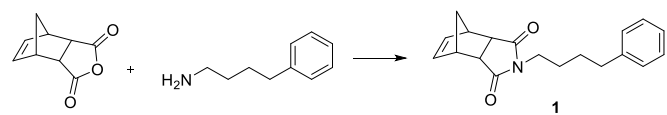

**Scheme 1**: Synthesis of monomer **1**

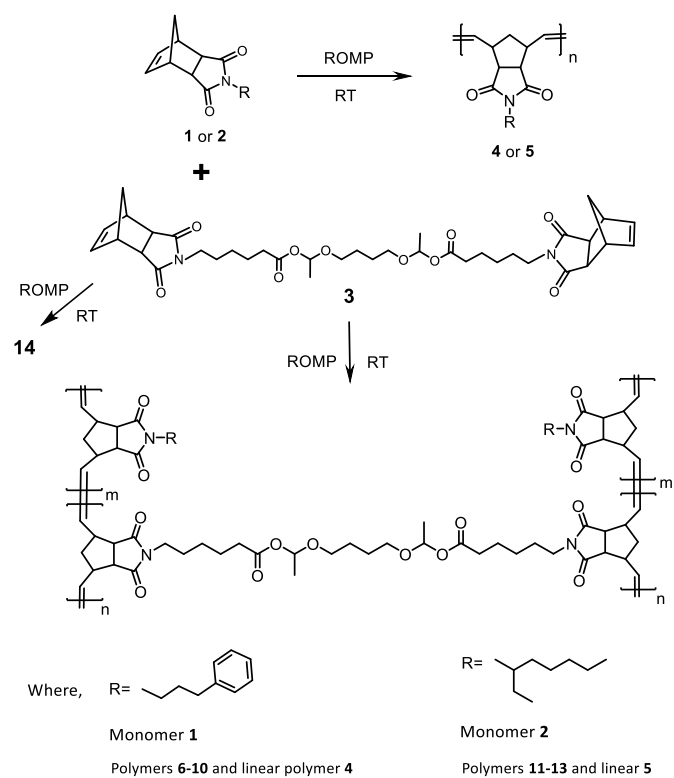

**Scheme 2**: Synthesis of linear polymers **4-5** and cross-linked polymers **6-14**

**Table 1**: Amount of monomer and initiator used in the synthesis of cross-linked materials **6-14** and calculated % Gel content

| Polymer | Difunctional monomer 3 % | 3 g, (mmol) | Monofunctional monomer 1 g, (mmol) | Monofunctional monomer 2 g, (mmol) | Initiator g, (mmol) | % Gel content |
|---|---|---|---|---|---|---|
| 6 | 5 | 0.042, (0.06) | 0.336 (1.14) | - | 0.02, (0.02) | 74 |
| 7 | 10 | 0.084, (0.12) | 0.318 (1.08) | - | 0.02, (0.02) | 85 |
| 8 | 25 | 0.208, (0.30) | 0.266 (0.90) | - | 0.02, (0.02) | 82 |
| 9 | 50 | 0.209, (0.30) | 0.088 (0.30) | - | 0.01, (0.01) | 81 |
| 10 | 75 | 0.313, (0.45) | 0.044, (0.15) | - | 0.01, (0.01) | 77 |
| 11 | 25 | 0.435, (0.63) | - | 0.516, (1.88) | 0.041, (0.05) | 71 |
| 12 | 50 | 0.870, (1.25) | - | 0.344, (1.25) | 0.041, (0.05) | 86 |
| 13 | 75 | 1.305, (1.88) | - | 0.172, (0.63) | 0.041, (0.05) | 82 |
| 14 | 100 | 0.422, (0.61) | - | - | 0.01, (0.01) | 83 |

Di-functional monomer **3** was synthesised and characterised following previous published procedures.[14] However, for the details of the synthesis and characterisation see Supporting Information and Figures S1 and S2.

### Synthesis of linear polymers 4 and 5

Monomers **1** and **2** were subjected to ROMP using Grubbs 1st generation ruthenium initiator to prepare linear polymers **4** and **5**, respectively (Scheme 2).

In a typical polymerisation, **1** (1.0 g, 3.4 mmol) was dissolved in dry DCM (2 mL) in a vial. In a separate vial containing a stirrer bar, ruthenium initiator (0.056 g, 0.068 mmol) was dissolved in dry DCM (0.5 mL). The monomer solution was added to the vial containing initiator and stirred for 12 h, at ambient temperature. The reaction mixture was then precipitated in hexane. The product was filtered and dried at

50 °C under reduced pressure to obtain **4** (yield 81%). Linear polymer **5** was prepared using the above method using monomer **2** (yield 85%).

### Preparation of cross-linked polymers 6-14

In a typical copolymerisation, monomers **3** (0.042 g, 0.06 mmol) and **1** (0.336 g, 1.14 mmol) were dissolved in dry DCM (3 mL) in a vial. In a separate vial containing a small stirrer bar, ruthenium initiator (0.02 g, 0.024 mmol) was dissolved in dry DCM (0.5 mL). The solution of the monomer was added to the solution of initiator and stirred until a gel was formed. Ethyl vinyl ether (10 fold excess with respect to initiator) was then added to the mixture to terminate the polymerisation. DCM was removed under reduced pressure giving cross-linked polymer **6**. Cross-linked polymers **7-10** were similarly synthesised.

Cross-linked polymers **11-13** were prepared using the above method but using monomers **2** and **3** (Table 1).

For the synthesis of cross-linked polymer **14**, monomer **3** (0.422 g, 0.606 mmol) was dissolved in dry DCM (2 mL) in a sample vial. Ruthenium initiator (0.01 g, 0.012 mmol) was dissolved in dry DCM (0.5 mL) in a separate sample vial containing a stirrer bar. The monomer solution was added to the initiator solution and stirring commenced at ambient temperature until a solid formed. DCM was then removed under reduced pressure resulting in cross-linked polymer **14**.

The gel fraction contents of cross-linked polymers **6-14** were determined by sol-gel extraction following previously published procedure (Table 1).[14]

### Acid-Catalysed Hydrolysis

Polymer disks (8 mm diameter, 1 mm thickness, ~80 mg) were prepared in a heat press at 130 °C and 10 ton. The degradation of cross-linked materials by acid-catalysed hydrolysis using bio-derived acetic (0.01 M, 30 mL, pH 3.3) and citric (0.01 M, 30 mL, pH 2.6) acids were investigated.

The acid hydrolysis process was repeated with HCl (1.6 M, 30 mL, pH 2.8).

In a typical test, the pre-weighed polymer disk was added to a round bottom flask containing a stirrer bar. Acid was added and stirred for 48 h at 50 °C or 144 h at ambient temperature. The solid was recovered by filtration, washed with pure water three times and dried under reduced pressure until a constant weight. The weight loss of the polymer disk was calculated by subtracting the polymer weight from the initial polymer weight divided by the initial polymer weight.

### Enzymatic Hydrolysis

Cross-linked polymers **6**, **7** and **8** containing 5%, 10% and 25% **3** respectively, were subjected to enzymatic degradation. The samples were incubated in potassium phosphate buffer (100 mM, pH 7.0) with *Thermobifida cellulosilytica* Thc_Cut1 (15 μM) at 50 °C on a rotary shaker at 100 rpm for 6 days. The samples were freeze dried then tested for solubility in DCM. As a control, cross-linked polymers **6-8** were incubated in buffer under the same conditions in the absence of enzyme.

### Rheological analysis

Rheological measurements were carried out using an AR-2000 rheometer (TA Instruments) with an 8 mm parallel plate. Polymers were compression moulded at 130 °C at 10 ton for 1 h to ensure the samples were stress free. Oscillatory frequency sweeps were performed in the range of 0.01 Hz to 100 Hz over temperatures from 100 °C to 190 °C. The oven was purged with nitrogen and a time weep was performed to show that there were no significant changes to the rheological properties of the sample over a time period of 6 h, which is significantly greater than the experimental run time. The results for each material were verified by performing repeat measurements on separate samples. The strain amplitude was varied to ensure the material was probed in the linear deformation regime and was typically between 0.5% and 2%. For each material the dynamic moduli from each temperature run were superimposed onto a linear master curve at 130 °C using Williams-Landel-Ferry (WLF) theory.[13] The data was analysed using RepTate software. The complex viscosity η* (Pa.s), was calculated using the equation below where G' is elastic modulus (Pa), G'' is viscous modulus (Pa) and ω is frequency (s$^{-1}$).[17]

$$\eta^* = \frac{\sqrt{G'^2 + G''^2}}{\omega}$$

## Results and discussion

### Polymer synthesis

Linear (**4-5**) and cross-linked (**6-14**) polymers were prepared via ROMP using Grubbs' 1st generation ruthenium initiator at ambient temperature. The thermosetting materials (**6-13**) were prepared by copolymerisation reaction of either **1** or **2**, and different amounts of cross-linking agent **3**, Scheme 2.

The resulting materials were anticipated to have a range of cross-linking densities, Table 1. Furthermore, monomer **3** was subjected to ROMP to prepare cross-linked material **14** with a maximum degree of cross-linking. Linear polymers **4** and **5** were prepared by ROMP of **1** and **2** respectively, to provide samples for comparative analysis. The gel content for cross-linked polymers **6-14** was found to be 71-86% indicating high efficiency of the ROMP cross-linking process (Table 1). The soluble parts of the cross-linked polymers were examined by $^1$H NMR and were found to contain some unreacted mono- and di-functional monomers.

### Degradation studies

**Acid-catalysed hydrolysis**. Insoluble cross-linked polymers **6-14** were subjected to acid-catalysed hydrolysis using bio-derived acids such as acetic and citric at ambient temperature and 50°C. The acid-catalysed reactions were also carried out in HCl to provide data for comparative analysis.

The same weight losses were observed, within experimental error, for the cross-linked polymers subjected to hydrolysis using bio-derived acetic and citric acids as well as HCl, upon which they became readily soluble in DCM, Figure 1. Moreover, polymers with a greater monomer **3** content have a higher cross-linking density and hence higher mass loss. The same weight loss trend was found for the acid-catalysed hydrolysis of cross-linked polymers **11-13**, (supporting information, Figure S3). The formation of soluble polymers was achieved after a time scale of 48 h at 50°C and 144 h, at ambient temperature. This indicates the efficiency of these three acids in hydrolysis for the transition from thermosets to thermoplastics.

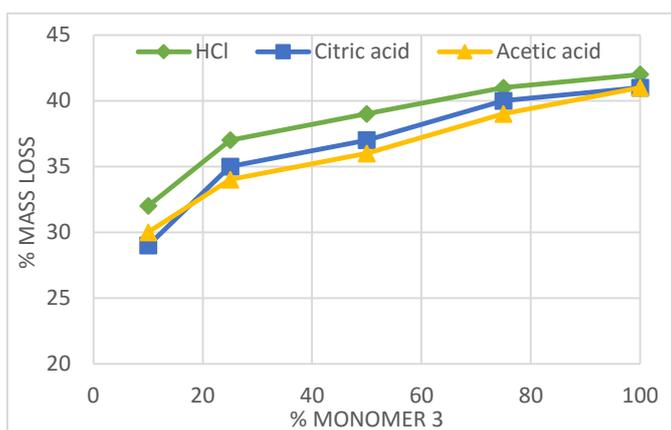

**Figure 1**: % Mass loss of cross-linked polymers **7-10** and **14** after acid-catalysed hydrolysis using citric, acetic and hydrochloric acid

**Soluble products**. The acetal ester linkage in cross-linked polymers **7-10** and **11-13** is cleaved upon acid catalysed hydrolysis resulting in the formation of the same soluble linear polymer products regardless of the level of cross-linking density and the only difference will be the pendant groups in the repeat units of monomers **1** and **2**. However, the $^1$HNMR spectra of the linear polymer products from **7-10** are more crucial in their structural identification as they contain phenyl rings in the pendent group of the repeat units of monomer **1** which can easily be identified. The $^1$HNMR spectra of the soluble polymer from cross-linked polymer 8 is shown in Figure 2 as a typical example of soluble linear polymers obtained from cross-linked polymers **7-10** upon acid-catalysed hydrolysis.

The $^1$H NMR spectrum of the soluble polymer obtained from the hydrolysis of cross-linked polymer **8** by acetic acid (Figure 2iii) shows the disappearance of resonances at 5.87 ppm, 1.36 ppm and 3.48 ppm attributing to methine proton (f), ethyl protons (h) and methylene protons (g), respectively, on the acetal ester linkage moiety (Figure 2i). Also, the aromatic protons (a-c) at 7.10-7.19 ppm, vinylic protons (d) at 5.67 ppm, and -CH$_2$-N- group at 3.40 ppm (e) can be seen which are identical to those observed for linear polymer **4** (Figure 2ii). The integration of methylene protons (e) and the aromatic protons (a-c) are the same for both polymer samples (Figures 2ii and 2iii). Furthermore, linear polymer **4** was subjected to acid-catalysed hydrolysis under the same conditions and only negligible weight loss was observed which is well within the experimental error of the measurements. This is indicative of the stability of the polymer backbone chain and the pendant groups in the repeat unit of monomer **1** during the acid-catalysed hydrolysis process. These observations clearly confirmed the successful hydrolysis of the acetal ester cross-linkages within the cross-linked polymers and the formation of soluble linear copolymer (supporting information, Figure S4). The formation of copolymer is further confirmed by the $^{13}$C NMR of soluble polymer resulting from cross-linked polymer **10** hydrolysed by HCl (supporting information, Figure S5) which shows the presence of carbonyl groups of the dicarboximides (a) at 178.6ppm, carboxylic acid (b) at 176.7pm and the aromatic carbons (c) at 132.0ppm. Similar results were obtained for acid-catalysed hydrolysis of cross-linked polymers **6-13** using citric acid and HCl.

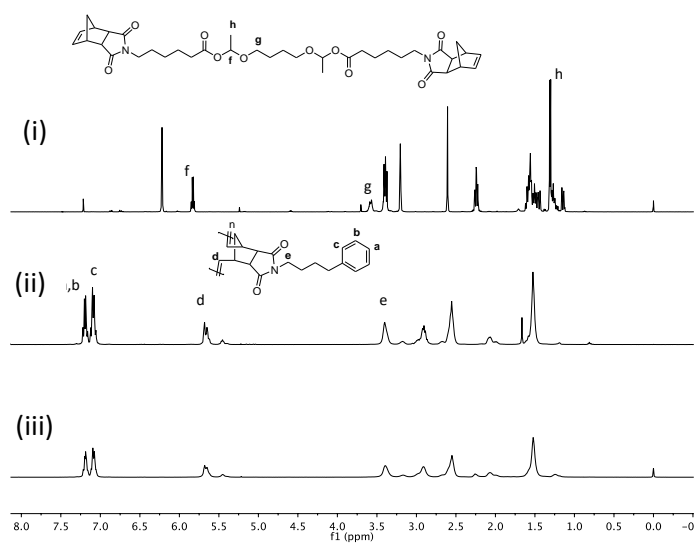

**Figure 2**: $^1$H NMR spectra in CDCl$_3$; (i) di-functional monomer **3,** (ii) linear polymer **4** and (iii) soluble polymer from cross-linked polymer **8** after acid-catalysed hydrolysis using acetic acid

Soluble polymer obtained from cross-linked polymer **7** after acid-catalysed hydrolysis using citric acid and HCl showed the same Tg, within experimental error, comparable to that of linear polymer **4** (Figure 3).

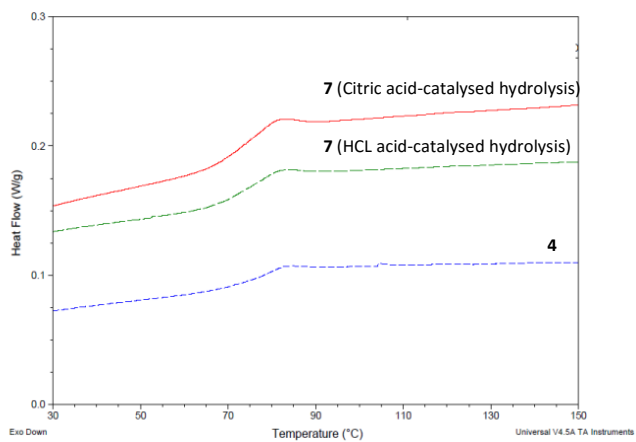

**Figure 3**: DSC traces showing the glass transition temperature for cross-linked polymer **7** after acid-catalysed hydrolysis with citric and HCl acids and linear polymer **4**

**Enzymatic hydrolysis**. The production of Thc_Cut1 in *E. coli* BL21-Gold(DE3) was successful, proven by SDS-PAGE analysis of the soluble and insoluble cell fractions where the strong protein bands just below 30 kDa corresponded well to the calculated mass of Thc_Cut1 (29.4 kDa). Thc_Cut1 were expressed intracellularly in its soluble and active form. High purity was achieved by purification over 6xHis-Tag. 40-60 mg purified enzyme were obtained from 400 mL cell culture. Common in cross-linked materials is the acetal ester linkage which goes through enzymatic degradation. Thermosetting materials with cross-linking density of 5-25% are the most interesting as far as the industrial application of the thermosets are concerned. Therefore, cross-linked polymers **6-8** containing 5, 10, and 25 %, respectively, difunctional monomer **3** were chosen for extensive enzymatic degradation study.

Insoluble cross-linked polymers **6-8** containing 5%, 10% and 25% **3**, respectively, were subjected to enzymatic degradation using cutinase (Thc_Cut1) upon which they became readily soluble in DCM. This confirms that Thc_Cut1 can effectively degrade the acetal-ester cross-links; transforming insoluble cross-linked ROMP thermosetting materials into linear, soluble polymers under mild and environmentally friendly conditions.

Furthermore, cross-linked polymers **6-8** were subjected to a blank test under the same conditions in the absence of enzyme, showing negligible weight losses.

The cross-linked polymers **9**-**14** show are expected to show the same behaviour as they all contain acetal ester linkage which goes through enzymatic degradation. The pendant groups in the repeat units of monomers **1** and **2** do not appear to influence the enzymatic degradation

**Soluble products**. In order to evaluate the products resulting from the enzymatic degradation of acetal ester linkage, the di-functional monomer **3** was first investigated by GC-MS, Figure 4. A compound with a mass of 211 was observed corresponding to **17** which was a consequence of compound **16**, formed via degradation but then going through a retro Diels-Alder reaction during GC-MS analysis, Scheme 3. Moreover, compound **15** with a mass of 178, was also identified through mono-silylation with BSTFA giving a mass of 234. The products of enzymatic degradation of cross-linked polymers **6**-**8** were then investigated by GC-MS. Similarly, the compound **15** with a mass of 178, due to degradation of the acetal ester linkage, was identified through mono-silylation with BSTFA giving a mass of 234, Scheme 3. However, this mass was not observed in control tests in the absence of enzyme. Moreover, mono-functional monomer **2** was subjected to enzymatic degradation in order to investigate the stability of the dicarboximide moiety, which was found to be enzymatically stable. The only product identified was compound **18** with a mass of 210 which is the consequence of monomer **2** undergoing a retro Diels-Alder reaction during GC-MS analysis, Scheme 3.

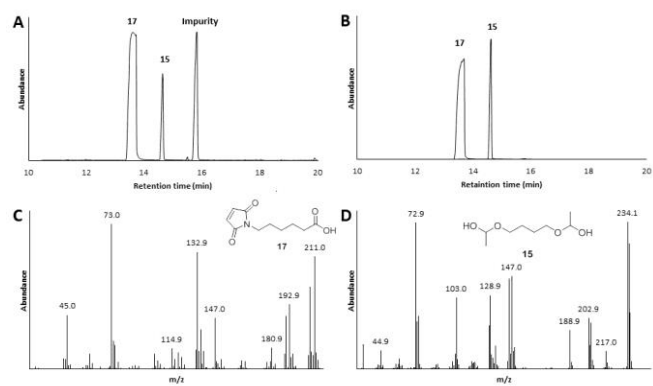

**Figure 4**: Analysis of the released products of enzymatic degradation of di-functional monomer **3**, determined by GC-MS, (A) chromatogram of expected hydrolysis products (B) extracted ion chromatogram for ion 234 and 211 (C) and (D) mass spectra of expected degradation products.

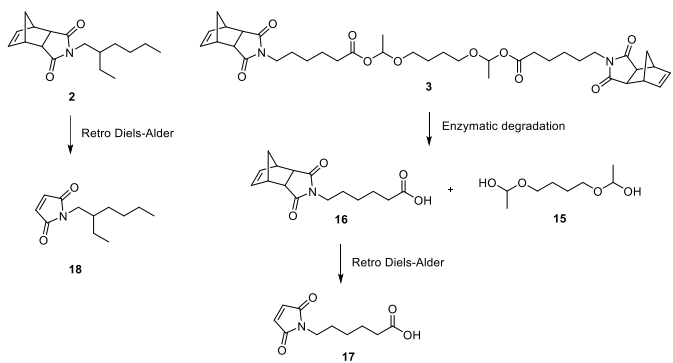

**Scheme 3**: Products of enzymatic degradation of monomers **2** and **3**, determined by GC-MS.

**Thermal Stability**

The thermal stability of the polymers **4**-**14** were investigated by TGA (Figure 5). Linear polymer **5** (containing no acetal ester groups) was shown to be stable to 350 °C and only a few percent weight loss was observed due to moisture contaminant or residual of solvent in the material. In contrast, cross-linked polymers **6**-**10** as well as **14** were shown to be stable to 150 °C, beyond which showed weight losses of 6%, 10%, 17%, 20%, 25% and 30%, respectively. The weight loss is due only to the thermal cleavage of the acetal ester linkages and is consistent with **6** and **14** having the lowest and the highest difunctional monomer content, respectively.

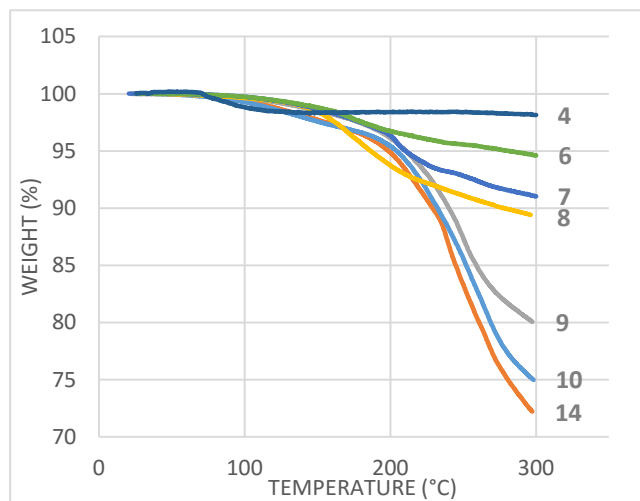

**Figure 5**: TGA thermographs observed during the thermal degradation of polymers **4**, **6-10** and **14**

## Rheological analysis

Rheological analysis has been shown to be significantly sensitive enough to discriminate between materials of linear and cross-linked architecture.[19,20,21] Particularly, it has been well established that linear oscillatory shear is sensitive to polymer molecular weight[22,23,24] and cross-linking densities in polymer networks.[18,25,26,27,28]

Cross-linked polymers **6** and **7** containing 5% and 10% **3**, respectively, were subjected to rheological analysis before and after acid-catalysed hydrolysis using acetic and citric. The complex viscosity for cross-linked polymers **6** and **7** are shown on the same scale before and after acid hydrolysis, Figure 6a and 6b respectively. The complex viscosities for the two hydrolysed samples are super-imposable indicating the successful hydrolysis of the acetal ester linkages and hence transformation into linear polymers.

Linear polymer **5**, cross-linked polymer **11** containing 25% **3**, before and after HCl acid hydrolysis were also studied, Figure 7. An average relaxation time of each sample is indicated by the reciprocal frequency at which the two moduli cross. Cross-linked polymer **11** has a longer relaxation time than its linear analogue **5**, which would be expected for a material with an increased mesostructure.

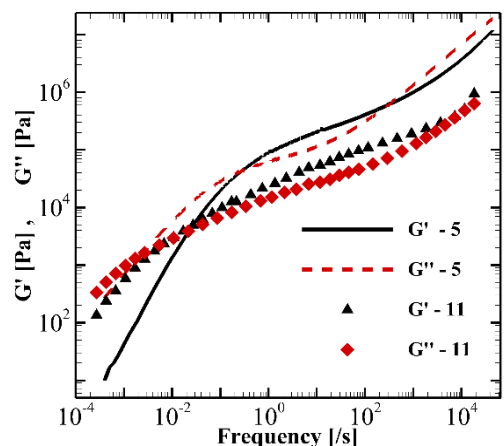

**Figure 7**: Dynamic moduli master curves for linear polymer **5** and cross-linked polymer **11**. The cross-over frequency when G' = G" is at a lower frequency for the cross-linked sample, indicating a much longer characteristic relaxation time.

The dynamic moduli clearly distinguish linear polymer **5** from cross-linked polymer **11**. The dynamic moduli for cross-linked polymers **11** and **12** containing 25% and 50% **3**, respectively, after HCl acid hydrolysis, are shown in Figure 8. The viscous and elastic moduli superimpose onto that of the linear polymer **5**. This confirms the acetal ester cross-links have been hydrolysed to produce linear polymers. Cross-linked polymer **12** was unmeasurable before hydrolysis due to the inability to mould the sample so that it was free of inhomogeneity. Furthermore, the complex viscosity for cross-linked polymers **11** and **12** after acid hydrolysis are compared to that of linear polymer **5** in Figure 9. At the limit of zero frequency, the zero shear viscosity for polymers **11** and **12** (700kPa.s) after acid hydrolysis is lower than that of the un-hydrolysed polymer **11** (1300kPa.s), which is expected after the break down of the acetal ester crosslinks.

A lower complex viscosity was observed in Figure 6 with hydrolysed polymers **6** and **7** (14kPa.s) compared to those obtained from hydrolysed **11** and **12** (700kPa.s) in Figure 9. The differing complex viscosities of these samples are attributed to the nature of the chemical structures in the pendent groups of monomers **1** and **2**, and the increased level of chain entanglement in hydrolysed **11** and **12**. In addition, rheological analysis showed that the dynamic moduli and complex viscosities for the linear polymer before and after acid treatment are superimposable, Figure 10. This indicates the stability of the linear polymers in acidic solutions.

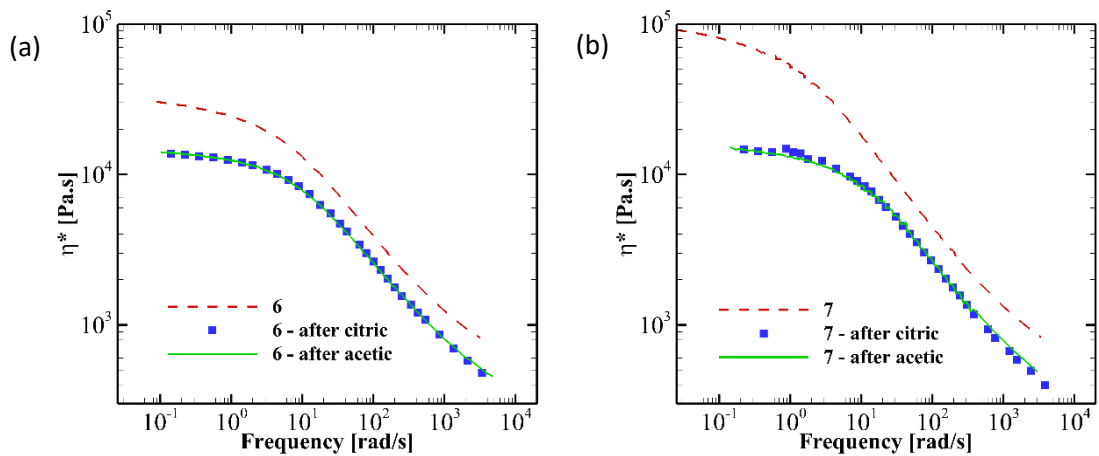

**Figure 6**: Complex viscosities for cross-linked polymers **6** (a) and **7** (b) before and after acid-catalysed

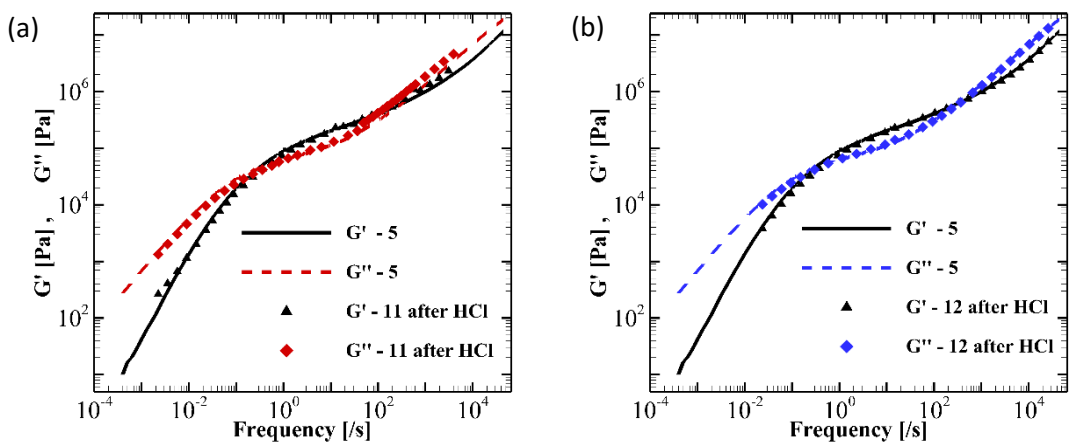

**Figure 8**: Dynamic moduli master curves for (a) linear polymer **5** with cross-linked polymer **11** after acid-catalysed hydrolysis using HCl and (b) linear polymer **5** with cross-linked polymer **12** after acid-catalysed hydrolysis using HCl

Rheological analysis clearly shows the effects of cross-linking density and acid hydrolysis on the complex viscosity and dynamic moduli of the polymers. Cross-linked polymers **11** and **12** after HCl acid hydrolysis showed the same rheological characteristics as that for linear polymer **5**. The effect of increased cross-link density on the rheological properties (complex viscosity and dynamic moduli) was also observable when comparing the cross-linked samples **6**, **7**, **11** and **12** before hydrolysis. The crucial finding here is that bio-derived acetic and citric acids were successful in fully hydrolysing the cross-links and forming linear polymers.

**Table 2**: Rheological properties of polymers studied

| Polymer | Relaxation time, $\tau$ s | Zero shear viscosity, $\eta^*_0$ Pa.s | Tg °C |
|---|---|---|---|
| 5 | 3.18 | 650,000 | 82.0 |
| 6 | 0.0355 | 31,300 | - |
| 6 (after acid) | 0.0211 | 14,000 | - |
| 7 | 0.234 | 93,300 | - |
| 7 (after acid) | 0.0235 | 14,600 | - |
| 11 | 117 | 1,350,000 | 83.1 |
| 11 (after acid) | 2.52 | 640,000 | 84.1 |
| 12 (after acid) | 2.67 | 573,000 | 84.7 |

Rheological temperature sweeps were performed on linear polymer **5** and cross-linked polymer **11** before and after hydrolysis using HCl (Table 2 and supporting information, Figure S6). The samples showed a similar glass transition temperature, Tg, of 83°C±2°C showing no difference between cross-linked or linear polymers. This can be explained due to the high degree of flexibility of the spacer in the crosslink moiety.[18]

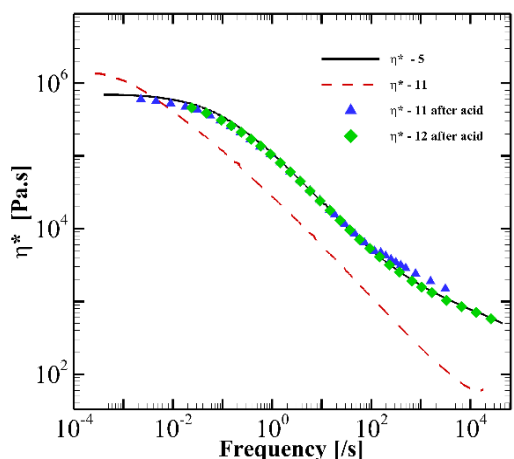

**Figure 9**: Complex viscosity for linear polymer **5**, cross-linked polymer **11** before and after acid-catalysed hydrolysis and cross-linked polymer **12** after acid-catalysed hydrolysis. Note that the curves for linear polymer **5** and hydrolysed polymers **11** and **12** superimpose. In the limit of low frequency, cross-linked polymer **11** has a higher viscosity than the other samples.

## Conclusions

Thermosetting polymers **6-14** with varying cross-link densities were synthesised via ROMP of mixtures of mono-functional monomers **1** and **2** and di-functional monomer **3** using Grubbs ruthenium initiator at ambient temperature. Furthermore, linear polymers **4** and **5** were prepared under the same reaction conditions to provide a comparison with the analysis of the cross-linked materials.

The cross-linked polymers were thermally stable to 150 °C, beyond which showed weight losses due to the thermal cleavage of the acetal ester linkages.

Cross-linked polymers **6-14** were subjected to acid-catalysed hydrolysis using bio-derived acetic and citric acid and compared with HCl. Furthermore, cross-linked polymers **6, 7** and **8** containing 5%, 10% and 25% **3**, respectively, were subjected to enzymatic degradation using cutinase from *Thermobifida cellulosilytica* (Thc_Cut1). All tested samples of insoluble cross-linked materials became readily soluble in DCM after acid hydrolysis and enzymatic degradation. However, it was not possible to compare the two degradation procedures due to small scale reactions and the difficulty of recovering the degraded products in the case of enzymatic degradation. As far as the duration is concerned the cross-linked materials degrade after 48 h for acid hydrolysis and 6 days for enzymatic degradation. The results of $^1$H NMR, rheological studies and enzymatic degradation confirm the breakdown of the acetal ester linkages allowing the transition from cross-linked to linear thermoplastic polymers. This is anticipated to facilitate the low temperature recycling of thermosetting materials, and the development of industrially important class of detachable/re-workable adhesives.

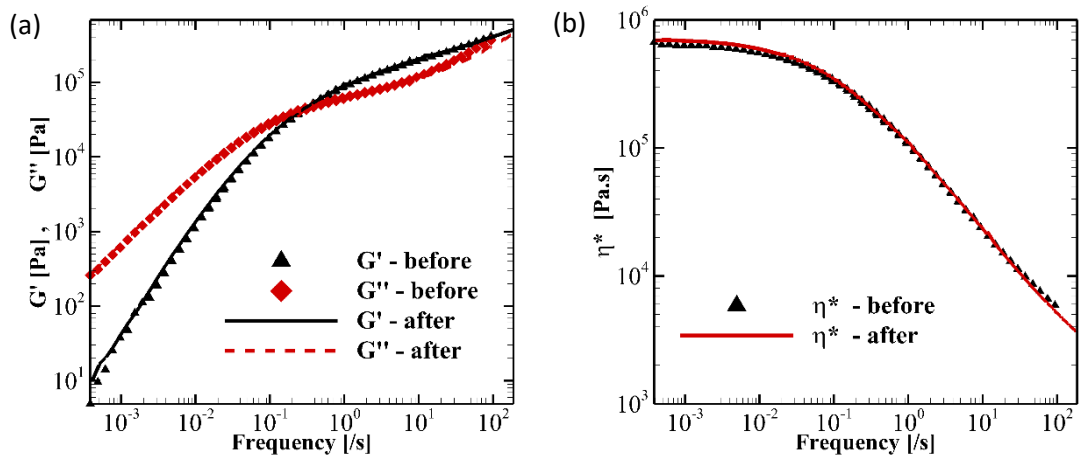

**Figure 10**: (a) Dynamic moduli and (b) complex viscosity for linear polymer **5** before and after acid-catalysed

## Acknowledgements

We acknowledge the EPSRC for Pathways-to-Impact award to DMH. We also acknowledge the support for KH, VP and GMG by the Federal Ministry of Science, Research and Economy (BMWFW), the Federal Ministry of Traffic, Innovation and Technology (bmvit), the Styrian Business Promotion Agency SFG, the Standortagentur Tirol, the Government of Lower Austria and Business Agency Vienna through the COMET-Funding Program managed by the Austrian Research Promotion Agency FFG.